\documentclass{pasj00}

\begin{document}

\SetRunningHead{Y. Takeda et al.}{N abundances and $^{12}$C/$^{13}$C ratios of red giants}
\Received{2019/05/17}
\Accepted{2019/08/21}

\title{Photospheric nitrogen abundances and \\ 
carbon $^{12}$C/$^{13}$C ratios of red giant stars
}

%

\author{
Yoichi \textsc{Takeda,}\altaffilmark{1,2}
Masashi \textsc{Omiya,}\altaffilmark{1,3}
Hiroki \textsc{Harakawa,}\altaffilmark{4,3} and
Bun'ei \textsc{Sato}\altaffilmark{5}

}

\altaffiltext{1}{National Astronomical Observatory, 2-21-1 Osawa, 
Mitaka, Tokyo 181-8588}
\email{takeda.yoichi@nao.ac.jp}
\altaffiltext{2}{SOKENDAI, The Graduate University for Advanced Studies, 
2-21-1 Osawa, Mitaka, Tokyo 181-8588}
\altaffiltext{3}{Astrobiology Center, National Institutes of Natural Sciences, 
2-21-1 Osawa, Mitaka, Tokyo 181-8588}
\altaffiltext{4}{Subaru Telescope, 650 N. A'ohoku Place, Hilo, HI 96720, U.S.A.}
\altaffiltext{5}{Tokyo Institute of Technology, 2-12-1 Ookayama, 
Meguro-ku, Tokyo 152-8550}

\KeyWords{stars: abundances --- stars: atmospheres ---  
stars: evolution --- stars: late-type 
}

\maketitle

\begin{abstract}
Nitrogen abundances and carbon isotope ratios ($^{12}$C/$^{13}$C) in the atmospheres
of red giants are known to be influenced by dredge-up of H-burning products and 
serve as useful probes to study the nature of evolution-induced envelope mixing.
We determined the [N/Fe] and $^{12}$C/$^{13}$C ratios for 239 late-G/early-K 
giant stars by applying the spectrum-fitting technique to the $^{12}$CN and 
$^{13}$CN lines in the $\sim$~8002--8005~\AA\ region, with an aim to investigate
how these quantities are related to other similar mixing-affected indicators
which were already reported in our previous work.
It was confirmed that [N/Fe] values are generally supersolar (typically by several 
tenths dex though widely differ from star to star), anti-correlated  
with [C/Fe], and correlated with [Na/Fe], as expected from theory.
As seen from their dependence upon stellar parameters, it appears that
mixing tends to be enhanced with an increase of stellar luminosity (or mass)
and rotational velocity, which is also reasonable from the theoretical viewpoint.
In contrast, the resulting $^{12}$C/$^{13}$C ratios turned out to be considerably
diversified in the range of $\sim$~5--50 (with a peak around $\sim 20$), without 
showing any systematic dependence upon C or N abundance anomalies caused by the 
mixing of CN-cycled material. It thus appears that our understanding on the 
photospheric $^{12}$C/$^{13}$C ratios in red giants is still incomplete, 
for which more observational studies would be required.
\end{abstract}

%


\section{Introduction}

Takeda, Sato, and Murata (2008, hereinafter referred to as Paper~I) conducted 
an extensive spectroscopic study on 322 targets (late-G through early-K giants) of 
Okayama Planet Search Program, which started at Okayama Astrophysical Observatory
since the beginning of this century and intended to search for planets around 
evolved red giants of intermediate mass by using the 188~cm reflector along with 
the newly installed High-Dispersion Echelle Spectrograph (HIDES). The purpose of 
Paper~I was to characterize the properties (stellar parameters and surface 
chemical abundances) of these program stars by analyzing their spectra. 

However, a puzzling result of appreciable oxygen deficiency was derived from the 
[O~{\sc i}] 5577 line (contradicting the prediction of standard stellar evolution 
calculations), which needed to be confirmed. Since no other oxygen lines 
were measurable on the spectra used in Paper~I (covering only $\sim$~5000--6200~\AA), 
we decided to reobserve many of these targets by using the updated HIDES (enabling 
3-times as wide wavelength coverage with 3 mosaicked CCDs) to obtain their spectra
covering longer wavelength region ($\sim$~5100--8800~\AA), where several 
important oxygen lines are available. 

Based on these new observational data for 239 stars collected in 2012--2013, 
Takeda et al. (2015, hereinafter referred to as Paper~II) redetermined their 
O abundances (with [O~{\sc i}] 6300/6363 as well as O~{\sc i} 7771--5 lines), 
and found that these new [O/H] (differential oxygen abundance relative to the Sun) 
results did not agree with those derived from [O~{\sc i}] 5577 in Paper~I.
A closer inspection further revealed that the reference solar abundance adopted 
in Paper~I was overestimated due to the neglect of C$_{2}$ molecular lines (which
are blended with the [O~{\sc i}] 5577 line), which should be the reason for the
appreciably subsolar [O/H]$_{5577}$. Actually, the characteristics of new O abundances 
established in Paper~II (as well as the C and Na abundances derived by non-LTE 
reanalysis of C~{\sc i} 5052/5380 and Na~{\sc i} 6160 lines) are consistent
with the theoretical prediction (i.e., appreciable deficiency in C, slight  
underabundance in O, and moderately excess in Na).  

Although the oxygen problem raised in Paper~I was settled as such and the abundance 
trends of C, O, and Na were shown to be by and large explained within the framework 
of canonical mixing theory (dredge-up of H-burning product), Paper~II could not
reach in-depth nature of evolution-induced abundance changes (e.g., how the abundance 
anomalies depend upon stellar parameters).
Besides, two important key indicators containing information on the nuclear-processed 
material salvaged from the interior still remain undetermined for the 239 stars 
studied in Paper~II; i.e., (i) N abundances and (ii) $^{12}$C/$^{13}$C ratios. 

It has been theoretically predicted and observationally reported since 
1960--1970s that red giants tend to show anomalies in the surface
abundances of N (increase) as well as $^{12}$C/$^{13}$C (lowering) 
as a result of contamination of CN-cycled product salvaged from 
the inner H-burning region caused by evolution-induced mixing.
However, since the details regarding how this mixing takes place 
are still uncertain, its clarification remains as an important task
in stellar astrophysics. Significant progress has recently been made in 
this field thanks to the extensive computer simulations based on the 
refined theory (including physically more realistic processes such as
thermohaline mixing or rotational mixing) as well as to the 
observational studies based on wealthy observational data covering 
wide range of objects (field giants, metal-poor giants, giants in clusters, 
etc.); see, e.g., Lagarde et al. (2019) and the references therein. 
Even so, since extensive spectroscopic studies targeting a large 
number of sample stars tend to be rather scarce still to date,
it may be worth reanalyzing the data of Paper~II in this context,
though our objects are limited to apparently bright nearby giants.

This situation motivated us to evaluate these two quantities based on the same 
observational material used in Paper~II, in order to complement our previous studies.
For this purpose, we decided to employ a group of $^{12}$CN and $^{13}$CN lines in 
the $\sim$~8002--8005~\AA\ region, which are known to be suitable and commonly used 
(e.g., Carberg et al. 2012; Sablowski et al. 2019). That is,, N abundances can be 
obtained with the help of the already known C abundances because absolute strengths 
of CN lines depend upon the product of C and N abundances, while $^{12}$C/$^{13}$C 
ratios are evaluated from the relative strengths of $^{12}$CN and $^{13}$CN lines. 

Accordingly, the aim of this paper is to present the results of these determinations, 
to discuss their trends in combination with the already established C, O, and Na 
abundances as well as various stellar parameters, and to compare them with theoretical 
calculations.

\section{Observational data and stellar parameters}

The 239 program stars in this study (and also in Paper~II) are 
subsamples of 322 stars in Paper~I, which are the targets of Okayama 
Planet Search Program. They are late G or early K giants (G5--K1~III) 
in the ranges of $\delta > -25^{\circ}$, $V < 6$, $0.6 < B-V < 1.0$, 
and $-3 < M_{V} < +2.5$, in which those cataloged as apparently 
variable stars or unresolvable binaries were excluded.
Most of them belong to the thin-disk population (only 5 are thick-disk 
stars; cf. Paper~I for the definition of stellar populations), and 19 stars 
are known to have substellar companions (taken from Paper~II; those designated 
as planet-host stars as of 2014).
Since all these program stars are apparently bright and limited to the solar
neighborhood (within several hundred parsec), 
very few stars are common with the targets of recent large survey projects intending 
very high-precision photometry from space (e.g., CoRoT --- Baglin et al. 2006; , 
{\it Kepler} --- Gilliland et al. 2010; K2 --- Howell et al. 2014; TESS --- 
Ricker et al. 2015) or studying the evolution of our Galaxy (e.g., APOGEE --- 
Majewski et al. 2017; {\it Gaia}-ESO --- Gilmore et al. 2012; LAMOST --- Cui et al. 2012; 
GALAH --- De Silva et al. 2015, etc.).
The observational data employed for this investigation (HIDES spectra with the
resolving power of $R \sim 67000$ covering the wavelength range of $\sim$~5100--8800~\AA) 
are the same as those used in Paper~II (see section~2 therein for more details). 
Regarding the solar spectrum used for the reference, we used the Moon spectrum included
in the spectrum database published by Takeda et al. (2005).   
Actually, only a narrow portion of 8001--8006~\AA\ (comprising the CN lines to be analyzed) 
is needed for this study, but this region is mildly contaminated by telluric water vapor 
lines. We removed them by dividing the raw spectrum of each star by the spectrum 
of $\alpha$~Leo (rapid rotator) by using the IRAF\footnote{
IRAF is distributed by the National Optical Astronomy Observatories,
which is operated by the Association of Universities for Research
in Astronomy, Inc. under cooperative agreement with the National 
Science Foundation.} task {\tt telluric}. 
A demonstrative example of this elimination process is depicted in figure~1a,
and the resulting telluric-removed spectra of two representative stars 
(HD~62509 and HD~27371) and the Moon are shown in figure~1b. 
The satisfactory level of this procedure differed from star to star, as the strengths 
of telluric lines were season-dependent and considerably diversified. Some feature 
often remained without being cleanly removed, which lead to a locally poor S/N ratio
(e.g., a weak hump at $\sim$~8005~\AA\ in the spectrum of HD~27371; cf. figure~1b).
Accordingly, depending on the stellar radial velocity, unfortunate cases occasionally 
happened where such defects appreciably influenced the spectral lines of our interest
(especially, weak $^{13}$CN lines at $\sim$~8004.7~\AA\ were apt to suffer this problem).

Regarding the atmospheric parameters [$T_{\rm eff}$ (effective temperature), 
$\log g$ (surface gravity), [Fe/H] (metallicity), $v_{\rm t}$ (microturbulence)]
and the corresponding model atmosphere for each star, we exclusively adopted those 
determined/used in Paper~I unchanged (as done in Paper~II). 
The same applies also for the relevant stellar parameters such as $v_{\rm e}\sin i$.
The only exception was $\log L$ (stellar luminosity),\footnote{
We do not explicitly discuss $M$ (stellar mass) in this paper, since the $M$ values
derived in Paper~I are likely to be appreciably overestimated for a number of red 
clump stars (constituting many of the 322 program stars in Paper~I) due to  
inappropriate application of a coarse grid of theoretical stellar evolutionary tracks  
(cf. Takeda \& Tajitsu 2015; Takeda et al. 2016). Nevertheless, if precise $M$ values
do not come to an issue, a rough correlation between $M$ and $L$ should still hold
(cf. Fig.~3a in Paper~I). In this sense, we may state that high $L$ stars tend to 
have larger $M$, and vice versa.} which were newly computed by using {\it Gaia}
DR2 parallaxes\footnote{Since Gaia DR2 parallaxes are not available for  
HD~3546, 45410, 62509, 147700, and 212430, we adopted the new Hipparcos reduction 
data (van Leeuwen 2007) for these 5 stars.}
 (Gaia Collaboration et al. 2016, 2018) instead of the Hipparcos 
parallaxes (ESA 1997) adopted in Paper~I, though the differences are generally 
insignificant as shown in figure~2a.
Figure 2b shows the $\log L$ vs. $T_{\rm eff}$ diagram (theoretical HR diagram), 
where our program stars are plotted along with the theoretical evolutionary 
tracks calculated by the PARSEC code (Bressan et al. 2012, 2013) for different 
mass values.
The mutual correlations of these parameters (which are summarized in ``tableE.dat''
available as the online material) are illustrated in figure~3. 

\section{Abundance determination}

\subsection{Synthetic spectrum fitting}

Our task was to analyze the $^{12}$CN and $^{13}$CN lines around $\sim$~8002--8005~\AA. 
The data of CN lines as well as atomic lines comprised in this region were taken 
from the list of Carlberg et al. (2012). Although we formally included the Fe~{\sc i} 
line at 8005.049~\AA\ according to Sablowski et al. (2019), its contribution turned out 
to be negligible. These spectral line data we adopted are summarized in table~1.

As done in Paper~II, we introduced a (depth-independent) factor $\phi_{\rm CN}$
by which the occupation numbers of CN molecules (computed from a model 
atmosphere with metallicity-scaled abundances of C and N) are to be multiplied 
to reproduce the observed  CN line strengths (see also subsection 3.3 in Takeda, 
Kawanomoto, \& Sadakane 1998). Likewise, $^{12}$C/$^{13}$C ratio should be
counted another free parameter necessary to match the spectral line features, 
because both $^{12}$CN and $^{13}$CN lines are involved. 
In addition, since  Fe~{\sc i} lines (especially Fe~{\sc i} 8002.576) show appreciable 
strengths in this region, Fe abundance ($A$(Fe))\footnote{
$A$(X) (logarithmic number abundance for element X, which is often written as 
$\log\epsilon_{\rm X}$) is defined in the usual normalization of $A$(H) = 12; i.e., 
$A$(X) $\equiv \log [N({\rm X})/N({\rm H})] + 12$.} 
is also to be adjusted.

The method of analysis is essentially the same as in Paper~II. Applying the numerical 
algorithm described in Takeda (1995), we required the best fit  between theoretical 
and observed spectra in the 8001-8006~\AA\ region while varying $\phi_{\rm CN}$, 
$^{12}$C/$^{13}$C, $A$(Fe), $v_{\rm M}$ (macrobroadening parameter),\footnote{
This $v_{\rm M}$ is the $e$-folding half-width of the Gaussian broadening 
function ($\propto \exp[-(v/v_{\rm M})^{2}]$), which represents the combined 
effects of instrumental broadening, macroturbulence, and rotational velocity
(cf. subsubsection 4.2.2 in Paper I).} and $\Delta \lambda$ (radial velocity 
or wavelength shift).\footnote{The abundances of other elements were fixed at the 
metallicity-scaled values.} How the theoretical spectrum for the converged 
solutions fits well with the observed spectrum for each star is 
displayed in figure~4.

While the $\phi_{\rm CN}$ solution converged successfully in almost all cases, 
determination of $^{12}$C/$^{13}$C turned out more difficult and delicate, because
$^{13}$CN line feature at $\sim$~8004.5~\AA\ is considerably weak and tends to  
suffer from spectrum defect (even if slight) caused by imperfect removal of
telluric lines (cf. section~2). Accordingly, we checked the appearance of fitting 
(especially for the $^{13}$CN feature) by eye, and grouped the results of 
$^{12}$C/$^{13}$C into four classes: (A) reliable (satisfactorily good fit; 115 stars), 
(B) less reliable (not necessarily satisfactory fit though acceptable; 76 stars), 
(C) unreliable (too poor fit to be tolerable; 23 stars), and (X) undetermined 
(solution not converged; 25 stars). Only the class-A and class-B solutions
determined for 191 stars were finally adopted in this study, while class-C 
ones were discarded. 

\subsection{Derivation of N abundances}

In the atmospheres of late G and early K giants under study ($T_{\rm eff} \gtsim 4500$~K), 
most of CNO are still in the stage of neutral atoms, while the fractions of 
molecules (such as CO) are insignificant. Under this condition, 
the number population of CN is practically regarded 
as being proportional to the product of the C and N abundances 
($\propto \epsilon_{\rm C}\epsilon_{\rm O}$),
Then, according to the definition of $\phi_{\rm CN}$, we may derive [N/Fe] 
(metallicity-scaled logarithmic nitrogen abundance relative to the Sun) as
\begin{equation}
[{\rm N}/{\rm Fe}] = [\phi_{\rm CN}] - [{\rm C}/{\rm Fe}],
\end{equation}
where  $[\phi_{\rm CN}] \equiv \log\phi_{\rm CN, star} - \log\phi_{\rm CN, \odot}$\footnote{
The value of solar $\log\phi_{\rm CN, \odot}$ resulting from the analysis of Moon spectrum
(cf. figure~1b) turned out to be $-0.15$~dex, which is not equal to zero (unlike expectation). 
This is because the absolute C and N abundances as well as the $gf$ values of 
CN lines adopted in the calculation were not perfectly adequate.}
and [C/Fe] was taken from Paper~I (mean result of similar high-excitation C~{\sc i} 
5052 and 5380 lines).

A remark may be due regarding errors involved in such derived N abundances.
As done in subsection~3.2 of Paper~II, we inversely calculated the equivalent width 
of the $^{12}$CN 8003.553 line (from the $\phi_{\rm CN}$ solution established by spectrum 
synthesis analysis), based on which the ambiguities of $\phi_{\rm CN}$ in response to
errors in atmospheric parameters were evaluated as summarized in table~2, where 
the similar parameter-dependences of C abundances (derived from C~{\sc i} 5380) 
also shown for comparison.
We can see from this table that the ambiguities in $T_{\rm eff}$ are most important
in abundance errors for both $\log\phi_{\rm CN}$ and $A$(C).
However, since the sense of abundance variation in response to changing $T_{\rm eff}$
is opposite with each other, the error in the resulting N abundance is further enhanced
($\sim 0.2$~dex for a typical $T_{\rm eff}$ uncertainty of 100~K; see the 3rd row in table~1).   
Accordingly, we should keep in mind that our N abundances may contain potentially larger 
errors than the abundances of other elements (C, O, and Na). 
As such, typical ambiguities would be $\ltsim$~0.1--0.2~dex for [C/Fe], [O/Fe], and [Na/Fe],
and $\ltsim$~0.2--0.3~dex for [N/Fe], though that for [C/N] may be as large as $\sim 0.3$.    

Meanwhile, general discussion on the errors of $^{12}$C/$^{13}$C ratio is difficult, 
because systematic errors closely related to the spectrum quality (which differs from 
star to star depending on the removal procedure of telluric lines; cf. section~2) 
are responsible for its reliability. We would roughly estimate that errors at least 
on the order of several tens percent may be involved even for the reliable class-A solutions. 

The final results of [N/Fe] and $^{12}$C/$^{13}$C derived for each star are summarized 
in tableE.dat (online material), where the Paper~II results of [C/Fe] (from C~{\sc i} 5052/5380), 
[O/Fe] (from O~{\sc i} 7771--5), and [Na/Fe] (from Na~{\sc i} 6161) as well as 
$A$(Li) (Takeda \& Tajitsu 2017) and $A$(Be) (Takeda \& Tajitsu 2014) are also presented.  

\section{Comparison with previous studies}

In figures 5 through 9 are compared the resulting [N/Fe] and $^{12}$C/$^{13}$C values 
(along with [C/Fe], [O/Fe], and atmospheric parameters given in ``tableE.dat'') with 
those derived for stars in common by previous representative studies: Lambert and Ries 
(1981) (figure~5), K{\ae}rgaard et al. (1982) (figure~6), Berdyugina (1993, 1994) 
(figure~7), Mishenina et al. (2006) (figure~8), and Tautvai\u{s}ien$\dot{{\rm e}}$ et al. 
(2010, 2013) (figure~9).

\subsection{Lambert and Ries (1981)}

Although appreciable discrepancies are observed between our atmospheric parameters and those 
adopted by Lambert and Ries (1981) (figures~5a--5c), the metallicity ([Fe/H]) and CNO 
abundances ([C/Fe], [N/Fe], and [O/Fe]) (figures~5d--5g) are more or less consistent with 
each other. Regarding $^{12}$C/$^{13}$C ratios, their results tend to be smaller as compared 
with our values (figure~5h). We suspect that their larger $v_{\rm t}$ (figure~4c) 
may be the cause, which acts in the direction of lowering $^{12}$C (though $^{13}$C being 
essentially unaffected) because $^{12}$CN lines are generally stronger while $^{13}$CN 
lines are considerably weak. For example, as to two representative stars shown in figure~1b, 
their $^{12}$C/$^{13}$C values are 16 (HD~62509) and 19 (HD~27371) (nearly the same), 
in contrast with our results of 15.6 (HD~62509) and 9.3 (HD~27371) (markedly different). 
This disagreement may be due to different choice of $v_{\rm t}$, since they adopted 
2.0~km~s$^{-1}$ (HD~62509) and 1.5~km~s$^{-1}$ (HD~27371), while our $v_{\rm t}$ values 
are 1.26~km~s$^{-1}$ (HD~62509) and 1.34~km~s$^{-1}$ (HD~27371).

\subsection{K{\ae}rgaard et al. (1982)}

The atmospheric parameters ($T_{\rm eff}$, $\log g$, [Fe/H]) of K{\ae}rgaard et al. (1982) 
are almost consistent with ours (figures~6a, 6b, and 6d), though they assumed 
$v_{\rm t}$ = 1.7 km~s$^{-1}$ (figure 6c).
Regarding [C/Fe], [N/Fe], and [O/Fe], although the correlation diagrams (figures~6e--6g) show 
appreciable scatters and agreement is not necessarily good, the general tendencies
of subsolar [C/Fe], supersolar [N/Fe], and almost near-solar [O/Fe] are similarly observed. 
They did not determine $^{12}$C/$^{13}$C ratios.

\subsection{Berdyugina (1993, 1994)}

Berdyugina's (1993, 1994) atmospheric parameters and CNO abundances are in reasonable
consistency with our results as can be recognized in figure~7, except that their $v_{\rm t}$ 
values tend to be somewhat higher (figure~7c). It is worth noting that a satisfactory 
agreement is seen also in the $^{12}$C/$^{13}$C ratios (figure~7h).

\subsection{Mishenina et al. (2006)}

Mishenina et al.'s (2006) atmospheric parameters ($T_{\rm eff}$, $\log g$, $v_{\rm t}$, 
and [Fe/H]) are in fairly good agreement with ours (figures~8a--8d).
Similar consistency is observed also for [C/Fe], [N/Fe], and [O/Fe] (figures~8e--8g),
though their [C/Fe] results tend to be slightly smaller than our values by 
$\sim$~0.1--0.2~dex (figure~8e). They did not determine $^{12}$C/$^{13}$C ratios.

\subsection{Tautvai\u{s}ien$\dot{{\rm e}}$ et al. (2010, 2013)}

The atmospheric parameters ($T_{\rm eff}$, $\log g$, $v_{\rm t}$, and [Fe/H]) of 
Tautvai\u{s}ien$\dot{{\rm e}}$ et al. (2010, 2013) are reasonably consistent with 
ours (figures~9a--9d). Regarding the abundance results, [O/Fe] and $^{12}$C/$^{13}$C 
ratios are in rough agreement (figures~9g and 9h). However, discrepancies are observed
in [C/Fe] as well as [N/Fe], in the sense that their values are almost constant at [C/Fe] 
$\sim -0.3$ and [N/Fe] $\sim +0.3$ despite that our values appreciably spread 
(cf. figures~9e and 9f).

\section{Discussion}

\subsection{Abundance trends and theoretical predictions}

We now discuss the results of [N/Fe] and $^{12}$C/$^{13}$ derived from our analysis,
where our attention is paid to the following points:
\begin{itemize}
\item
Do they show any significant dependence upon the stellar parameters; i.e., 
$T_{\rm eff}$, $L$, [Fe/H], and $v_{\rm e}\sin i$?
\item
How do they correlate with the abundances of other elements (Li, Be, C, O, and Na)
which are also likely to be influenced by evolution-induced envelope mixing? 
\item
Are the observed trends consistent with the predictions from recent stellar evolution
calculations?
\end{itemize}

In preparation for discussing these issues, how the resulting [N/Fe] and $^{12}$C/$^{13}$C 
are correlated with the relevant stellar parameters or light-element abundances is 
illustrated in figure~10 and figure~11, respectively.
Regarding the theoretical comparison, we invoked (as in Paper~II) Lagarde et al.'s (2012) 
extensive simulations, where they adopted two kinds of mixing treatments: (i) only 
conventional mixing and (ii) conventional mixing plus rotational and thermohaline mixing.
The predicted surface abundance changes (computed for $z=0.004$ and $z=0.014$ with 
three initial masses of 1.5, 2.5, and 4~$M_{\odot}$) are plotted against $T_{\rm eff}$ in 
figures 12a,a$'$ (N abundances) and 12b,b$'$ ($^{12}$C/$^{13}$C), and their mutual correlations 
are displayed in figures~12c,c$'$. We can see from these figures that the abundance anomalies
(i.e., enrichment in N and decrease in $^{12}$C/$^{13}$C) tend to be enhanced for
larger $M$ and by including rotational/thermohaline mixing (as expected), 
while the results do not depend much upon the metallicity. 
In figure~13 are plotted the observed trends of 
N-to-C abundance ratios (important indicator of abundance anomaly due to mixing of 
CN-cycled material) against [Fe/H] (figure~13a), $^{12}$C/$^{13}$C (figure~13b),
and [O/C] (figure~13c, where the theoretical curves are also overplotted).

\subsection{Nitrogen abundances}

We can confirm that the [N/Fe] values show dependences upon the abundances 
of C, O, and Na. That is, [N/Fe] is anti-correlated with [C/Fe] (figure~10g) 
as well as [O/Fe] (figure~10h), while correlated with [Na/Fe] (figure~10i).
These trends are consistent with theoretical predictions (cf. Fig.~11 in Paper~II,
figure~12a,a$'$), which can be interpreted as the first dredge-up of CN-cycle and NeNa-cycle
products (i.e., deficiency of C, enhancement of N and Na, marginal decrease of O).
It can be seen from figure~13a that the systematic trend of increasing (negative) 
[C/N] toward $\sim 0$ with a decrease in [Fe/H] is consistent with Fig.~6
of Lagarde et al. (2019), who recently carried out an extensive theoretical study 
of C and N abundances for giant stars using {\it Gaia}--ESO survey data.
Likewise, figure~13a shows that the [N/C] is positively correlated with [O/C],
and this tendency is in agreement with the theoretical prediction.
This trend is just the same as what Takeda, Jeong, and Han (2019) found in their 
CNO abundance study of Hertzsprung-gap stars (see Fig.~13h therein).
Accordingly, we may state that relative variations (or correlations) of C, N, O, 
and Na abundances are well explained by the canonical stellar evolution theory.

Regarding the correlation with other stellar parameters, we can see that [N/Fe] 
tends to increase with [Fe/H] (figure~10c), as already mentioned above in 
connection with figure~13a. Also recognized are the increasing trends 
of [N/Fe] for an increase in $T_{\rm eff}$ (figure~10a), $L$ (figure~10b), and
$v_{\rm e}\sin i$ (figure~10d), the interpretation of which is not straightforward 
as these three parameters are mutually correlated (cf. figures~3a, 3d, and 3e). 
While the $T_{\rm eff}$-dependence is probably nothing but superficial caused by 
the $v_{\rm e}\sin i$ vs. $T_{\rm eff}$ correlation, it would be reasonable 
to regard that both $v_{\rm e}\sin i$ and $L$ can directly affect the efficiency of 
mixing, because (i) higher rotation is generally known to enhance mixing
in red giants and (ii) higher $L$ generally means larger $M$ (cf. footnote~2) 
which tends to cause larger anomaly as theoretically expected (figure~12a,a$'$).
Yet, it is difficult to discriminate these two factors, as $v_{\rm e}\sin i$ 
and $L$ are related with each other (figure~3e).

Meanwhile, we should be cautious about the positive correlations in [N/Fe] with $A$(Li) 
(figure~10e) and $A$(Be) (figure~10f), which contradict the naive expectation 
of anti-correlation (because envelope mixing should act in the direction of 
destroying such vulnerable species as Li and Be). Here, their dependence upon 
other stellar parameters should be taken into consideration: $A$(Li) depends upon 
$T_{\rm eff}$ (decreasing with a lowering of $T_{\rm eff}$; cf. Fig.~19d in 
Takeda \& Tajitsu 2017) while $A$(Be) tends to increase  with [Fe/H] (cf. Fig.~5d 
in Takeda \& Tajitsu 2014). Accordingly, we suspect that these apparently systematic 
trends seen in [N/Fe] vs. $A$(Li) as well as [N/Fe] vs. $A$(Be) relations
simply reflect the parameter-dependence of [N/Fe] (increasing with $T_{\rm eff}$ 
as well as [Fe/H]) mentioned above.

\subsection{Carbon isotope ratios}

Regarding the carbon isotope ratio, our $^{12}$C/$^{13}$C values widely distribute 
between $\sim$~5 and $\sim$~50, where many of them are in the range of $\sim$~10--30 with 
a peak around $\sim 20$, since the mean value $\langle^{12}$C/$^{13}$C$\rangle$ 
($\pm \sigma$: standard deviation) is 17.9 ($\pm 7.5$) (for 115 class-A values) and 
18.2 ($\pm 9.6$) (for 191 class-A + class-B values).
Unlike the case of [N/Fe], meaningful trends of $^{12}$C/$^{13}$C in comparison with
other abundances or stellar parameters are barely observed in figure~11, except for 
weak positive correlations with [Fe/H] (figure~11c), $A$(Li) (figure~11e), and 
$A$(Be) (figure~11f). While we should be careful in interpreting these apparent tendencies 
in terms of metallicity and Li/Be abundances for the same reason as the case 
of [N/Fe] (see the last paragraph in subsection~5.1), we note that 
Lambert, Domity, and Sivertsen (1980) also reported similar correlation between 
$^{12}$C/$^{13}$C and $A$(Li) (cf. their Fig.~9 therein).

It was rather unexpected that we could not find in $^{12}$C/$^{13}$C any sign of 
systematic dependence upon the abundance anomaly caused by the mixing of CN-cycled 
product (figure~11g, 11h, 11i). Actually, as seen from $^{12}$C/$^{13}$C vs. 
[N/Fe] (figure~12c,c$'$) and $^{12}$C/$^{13}$C vs. [C/N] (figure~13b) diagrams, 
an especially large scatter of $^{12}$C/$^{13}$C ranging from $\sim 5$ to $\sim 50$
is observed for stars showing appreciable CN anomalies ([N/Fe]~$\sim$~+0.4 or 
[C/N]~$\sim -0.6$). As such, our observational data are in marked conflict with
theoretical predictions (figure~12c,c$'$). How could it be possible to realize 
such a high $^{12}$C/$^{13}$C ratio of $\ltsim 50$ (i.e., less $^{13}$C indicating 
less contamination of H-burning material) and high N abundance (indicative of 
efficient dredge-up) simultaneously? If this large dispersion of $^{12}$C/$^{13}$C 
at a given [N/Fe] or [C/N] is real, this isotope ratio may be controlled by 
(not only the extent of mixed CN-cycled gas but also) some other unknown mechanism. 

Comparing Lagarde et al.'s (2019) Fig.~11 with our figure~13b, we can see that 
the observational $^{12}$C/$^{13}$C data of their reference stars (cluster stars 
as well as field stars) are quite uniform around $\sim$~10--20 irrespective of [C/N] 
(being more or less consistent with their theoretical prediction including thermohaline 
instability effect),
 which is different from our results and those of previous
studies (e.g., Lambert \& Ries 1981; Berdyugina 1993, 1994)
reporting that some field red giants show  
large $^{12}$C/$~{13}$C ratios (up to $\sim 50$ or even more). 
The reason for this distinction (i.e., why higher $^{12}$C/$^{13}$C stars 
are not seen in Lagarde et al.'s sample) is not clear, which might be
due to different constituent in terms of stellar evolutionary status
(i.e., number ratio of stars in the red clump to those ascending 
the giant branch).

To be fair, however, we can not rule out a possibility that our large $^{12}$C/$^{13}$C 
values might suffer appreciable systematic errors (which are difficult to estimate; 
cf. subsection~3.2), since its determination becomes progressively more difficult as 
this ratio increases (because of the considerably weakened $^{13}$CN line strength). 
At present, what we can do is to simply present our observational results,
the validity of which is hopefully to be checked by further investigations. 
In any event, more extensive and precise spectroscopic determinations of $^{12}$C/$^{13}$C 
ratios for many red giants would be desired in order to understand their behaviors in 
context of other mixing-induced abundance anomalies and theoretical calculations.  

\section{Summary and conclusion}

\begin{itemize}
\item
In our recent work (Takeda et al. 2015), the photospheric abundances of 
C, O, and Na for 239 late-G and early-K giants (targets of the Okayama 
Planet Search Program) were studied with an aim to investigate their 
mixing-induced anomalies in comparison with theoretical predictions 
from stellar evolution calculations.
\item
In order to supplement the previous study, we focused in this paper 
on deriving their N abundances and $^{12}$C/$^{13}$C ratios, which 
are also useful probes to study the mixing of H-burning products 
salvaged from the interior.
\item
For this purpose, we applied a spectrum-fitting analysis to a group of CN 
lines existing in the $\sim$~8002--8005~\AA\ region, by which N abundances 
could be determined with the help of the already known C abundances, while 
$^{12}$C/$^{13}$C ratios were derived from the relative strengths of 
$^{12}$CN and $^{13}$CN lines. 
\item
The resulting [N/Fe] and $^{12}$C/$^{13}$C values (along with the C and O 
abundances) were compared with those published in four representative papers, 
and a reasonable consistency was confirmed regarding the general feature of 
CNO abundance anomalies (under/over-abundance of C/N, marginal deficiency 
of O) and the trend of $^{12}$C/$^{13}$C ratios ($\sim 10$ to several tens). 
\item
It was confirmed that the photospheric [N/Fe] values of these giants are 
generally supersolar typically by several tenths dex (though widely 
differ from star to star), anti-correlated with [C/Fe], and correlated 
with [Na/Fe]. These trends of abundance anomalies are consistent with the 
theoretical expectations, which suggests that mixing-related abundance 
variations of these light elements are reasonably explained by recent stellar 
evolution calculations. 
\item
As seen from the dependence of [N/Fe] upon stellar parameters, we may 
state that mixing tends to be enhanced with an increase of stellar luminosity 
(or mass) as well as rotational velocity, which is reasonable also from 
the theoretical point of view.
\item
In contrast, the resulting $^{12}$C/$^{13}$C ratios turned out to be considerably
diversified in the range of $\sim$~5--50 (with a peak around $\sim 20$) without 
showing any systematic dependence upon abundance anomalies caused by the 
mixing of CN-cycled material, which apparently disagrees with theoretical 
predictions. It thus appears that our understanding on the nature of photospheric 
$^{12}$C/$^{13}$C ratios in red giants is still incomplete, for which more 
observational studies would be required.
\end{itemize}

\bigskip

This research has made use of the SIMBAD database, operated by CDS, 
Strasbourg, France. This work has also made use of data from the European Space 
Agency (ESA) mission {\it Gaia} (https://www.cosmos.esa.int/gaia), processed 
by the {\it Gaia} Data Processing and Analysis Consortium (DPAC,
https://www.cosmos.esa.int/web/gaia/dpac/consortium). Funding for the DPAC
has been provided by national institutions, in particular the institutions
participating in the {\it Gaia} Multilateral Agreement.


\onecolumn 

\newpage

\setcounter{table}{0}
\begin{table*}[h]
\caption{Adopted data of spectral lines.}
\begin{center}
\footnotesize
\begin{tabular}{ccccc}\hline\hline
Species & $\lambda_{\rm air}$ & $\chi_{\rm low}$ & $\log gf$ & Source \\
        & ($\rm\AA$) & (eV) & (dex) &  \\
\hline
Fe~{\sc i} & 7998.944 & 4.371 & +0.1489 & CCSM\\
$^{12}$CN & 7999.214 & 1.400 & $-$2.0287 & CCSM\\
$^{12}$CN & 7999.214 & 1.600 & $-$1.8041 & CCSM\\
$^{13}$CN & 7999.408 & 0.090 & $-$1.6383 & CCSM\\
$^{13}$CN & 7999.460 & 1.460 & $-$1.2644 & CCSM\\
$^{13}$CN & 7999.465 & 0.220 & $-$1.7282 & CCSM\\
$^{12}$CN & 7999.846 & 0.100 & $-$1.9830 & CCSM\\
$^{12}$CN & 8000.261 & 0.190 & $-$1.4962 & CCSM\\
$^{12}$CN & 8000.316 & 1.470 & $-$1.6091 & CCSM\\
Nd~{\sc ii} & 8000.757 & 1.091 & $-$1.2220 & CCSM\\
$^{13}$CN & 8001.369 & 0.030 & $-$2.8962 & CCSM\\
$^{12}$CN & 8001.524 & 1.420 & $-$1.6253 & CCSM\\
$^{12}$CN & 8001.652 & 1.480 & $-$1.6091 & CCSM\\
$^{13}$CN & 8002.214 & 0.050 & $-$2.1805 & CCSM\\
$^{13}$CN & 8002.367 & 1.490 & $-$1.8327 & CCSM\\
$^{12}$CN & 8002.412 & 0.180 & $-$1.4962 & CCSM\\
$^{13}$CN & 8002.571 & 0.210 & $-$1.7212 & CCSM\\
Fe~{\sc i} & 8002.576 & 4.580 & $-$2.2400 & CCSM\\
Al~{\sc i} & 8003.185 & 4.087 & $-$1.8791 & CCSM\\
$^{12}$CN & 8003.213 & 0.120 & $-$1.9431 & CCSM\\
Fe~{\sc i} & 8003.227 & 5.539 & $-$2.3889 & CCSM\\
$^{13}$CN & 8003.311 & 1.340 & $-$2.0883 & CCSM\\
Ti~{\sc i} & 8003.485 & 3.724 & $-$0.2000 & CCSM\\
$^{12}$CN & 8003.553 & 0.310 & $-$1.6440 & CCSM\\
$^{12}$CN & 8003.910 & 0.330 & $-$1.6478 & CCSM\\
$^{12}$CN & 8004.036 & 0.060 & $-$2.9245 & CCSM\\
$^{13}$CN & 8004.550 & 0.120 & $-$1.5918 & CCSM\\
$^{13}$CN & 8004.715 & 0.070 & $-$2.0814 & CCSM\\
$^{13}$CN & 8004.801 & 0.100 & $-$1.6144 & CCSM\\
Fe~{\sc i} & 8005.049 & 5.587 & $-$5.5180 & SJIS\\
Zr~{\sc i} & 8005.248 & 0.623 & $-$2.1901 & CCSM\\
$^{13}$CN & 8006.065 & 1.410 & $-$1.6517 & CCSM\\
$^{13}$CN & 8006.126 & 0.240 & $-$1.7122 & CCSM\\
Si~{\sc i} & 8006.459 & 6.261 & $-$1.7231 & CCSM\\
Fe~{\sc i} & 8006.703 & 5.067 & $-$2.1280 & CCSM\\
$^{12}$CN & 8006.925 & 1.600 & $-$1.7878 & CCSM\\
$^{13}$CN & 8007.211 & 1.480 & $-$1.2652 & CCSM\\
Co~{\sc i} & 8007.242 & 4.146 & +0.1159 & CCSM\\
$^{12}$CN & 8007.582 & 0.110 & $-$1.9586 & CCSM\\
$^{13}$CN & 8007.882 & 0.030 & $-$2.8962 & CCSM\\
$^{13}$CN & 8007.904 & 0.050 & $-$2.1415 & CCSM\\
Si~{\sc i} & 8008.387 & 6.079 & $-$1.8289 & CCSM\\
\hline
\end{tabular}
\end{center}
\footnotesize
Abbreviation code for the source of $gf$ values:
CCSM --- Carlberg et al. (2012), SJIS --- Sablowski et al. (2019).
\end{table*}

\setcounter{table}{1}
\begin{table*}[h]
\caption{Abundance variations in response to changing atmospheric parameters.}
\begin{center}
\footnotesize
\begin{tabular}{ccccccc}\hline\hline
Line                  & $\Delta_{T+}$    & $\Delta_{T-}$    & $\Delta_{g+}$    & $\Delta_{g-}$    & $\Delta_{v+}$     & $\Delta_{v-}$    \\
\hline
 CN~{\sc i} 8003      & $+0.089$ (0.030) & $-0.066$ (0.034) & $+0.029$ (0.014) & $-0.025$ (0.014) & $-0.007$ (0.003)  & $+0.008$ (0.004) \\
 C~{\sc i} 5380       & $-0.088$ (0.011) & $+0.099$ (0.014) & $+0.081$ (0.004) & $-0.080$ (0.004) & $-0.003$ (0.002)  & $+0.002$ (0.002) \\
 N ($\equiv$~CN~/~C)  & $+0.176$ (0.022) & $-0.164$ (0.023) & $-0.052$ (0.012) & $+0.055$ (0.012) & $-0.004$ (0.003)  & $+0.006$ (0.004) \\
 O~{\sc i} 7774       & $-0.166$ (0.012) & $+0.182$ (0.015) & $+0.090$ (0.006) & $-0.091$ (0.005) & $-0.021$ (0.008)  & $+0.020$ (0.008) \\
Na~{\sc i} 6160       & $+0.076$ (0.006) & $-0.079$ (0.005) & $-0.012$ (0.003) & $+0.011$ (0.002) & $-0.048$ (0.012)  & $+0.050$ (0.012) \\
\hline
\end{tabular}
\end{center}
\footnotesize
Changes of the logarithmic abundances (expressed in dex) derived from each line 
in response to varying $T_{\rm eff}$ by $\pm 100$~K, $\log g$ by $\pm 0.2$~dex,
and $v_{\rm t}$ by $\pm 0.2$~km~s$^{-1}$. Shown are the mean values averaged over
each of the stars, while those in parentheses are the standard deviations.
The results for O~{\sc i} 7774 are taken from Table~3 of Paper~II.
\end{table*}

\newpage

\setcounter{figure}{0}
\begin{figure*}[p]
  \begin{center}
    \FigureFile(90mm,160mm){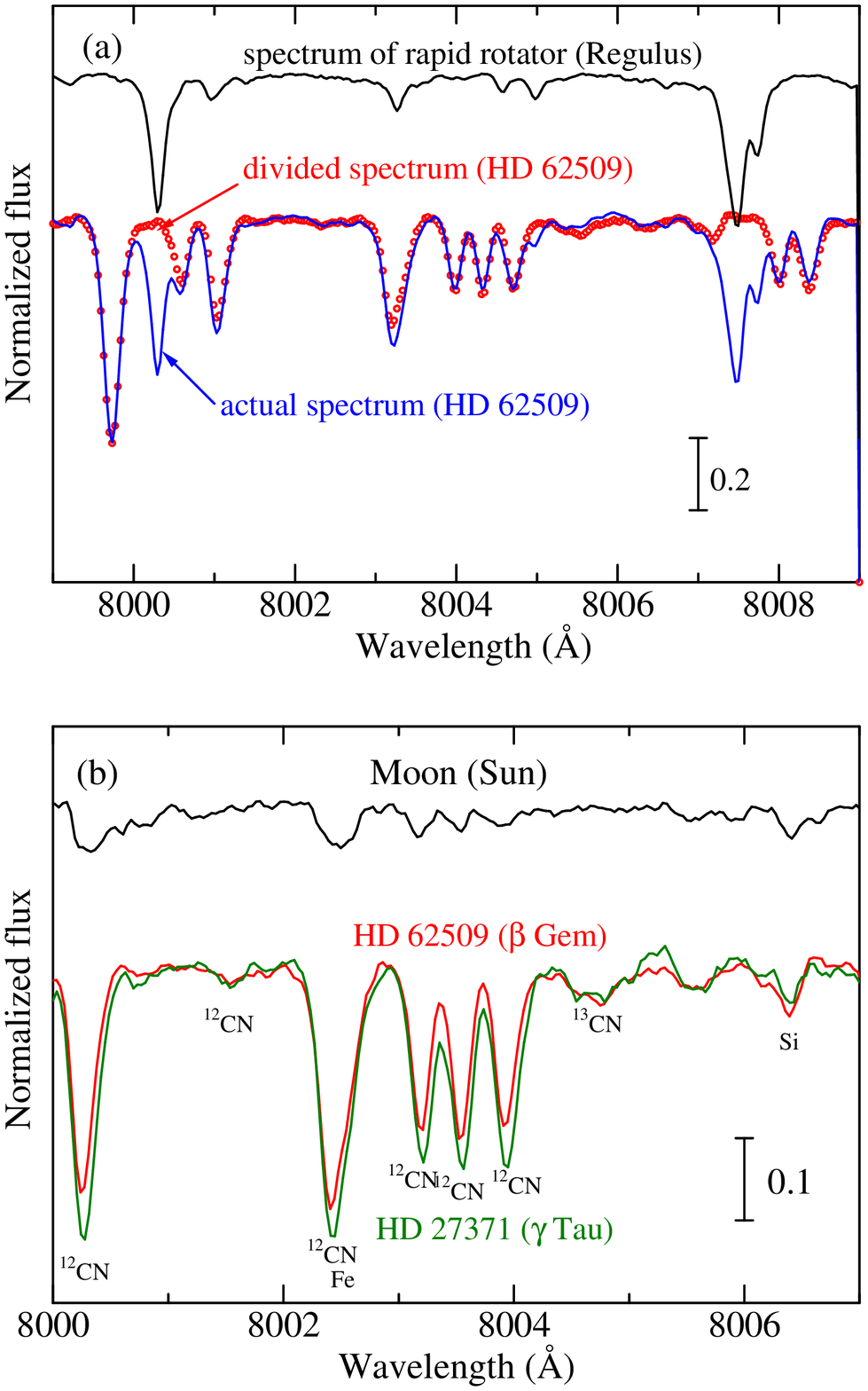}
  \end{center}
\caption{
(a) Example of how the telluric lines (due to H$_{2}$O vapor) are removed 
in the 7999--8009~\AA\ region, shown for the representative case of HD~62509. 
Dividing the actual stellar spectrum (blue line) by the 
spectrum of a rapid rotator (Regulus, black line) results in the final 
spectrum (red open circles). Spectra are shown in the raw 
wavelength scale without any radial-velocity correction. 
(b) Comparison of the telluric-removed spectra (in the 8000--8007~\AA\ 
region) of the Moon (Sun, black line), HD~62509 (red line), and HD~27371 
(green line). Radial velocity shifts of these spectra are so corrected 
that the wavelengths of stellar lines correspond to the laboratory values.  
}
\end{figure*}

\setcounter{figure}{1}
\begin{figure*}[p]
  \begin{center}
    \FigureFile(80mm,150mm){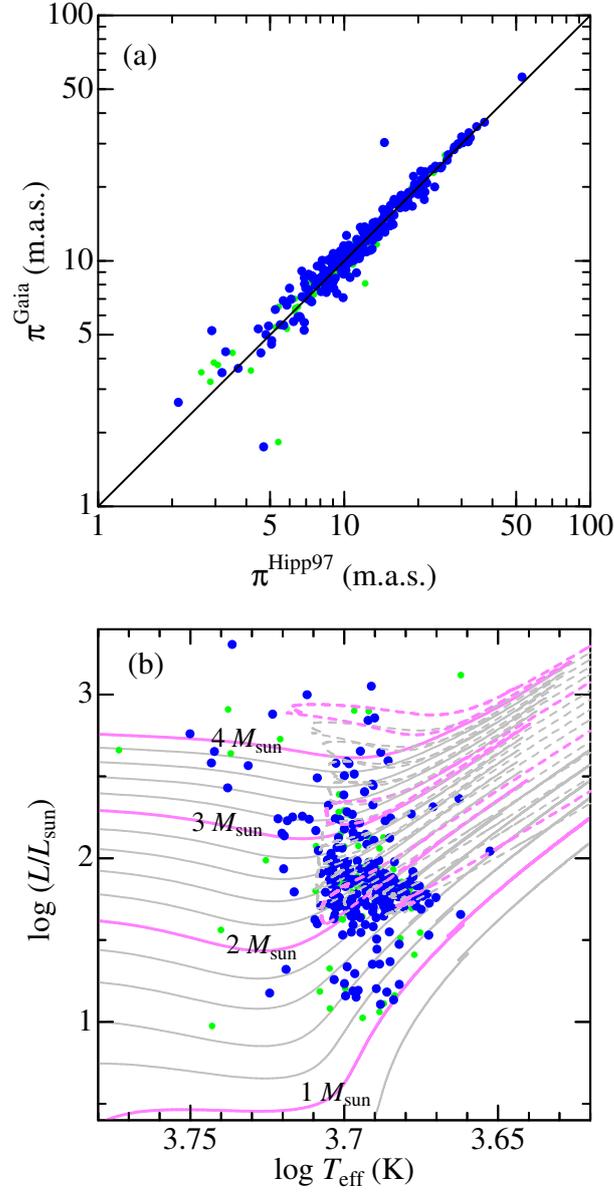}
  \end{center}
\caption{
(a) Comparison of the Hipparcos parallaxes (ESA 1997) used in Paper~I 
for evaluating stellar luminosities with the Gaia DR2 parallaxes 
(Gaia Collaboration et al. 2016, 2018) adopted in this paper. 
Our sample stars are denoted by larger (blue) symbols, while those stars 
studied only in Paper~I (but not included in this study as well as in Paper~II) 
are by smaller (green) ones.   
(b) Theoretical evolutionary tracks illustrated on the 
$\log T_{\rm eff}$--$\log L/{\rm L}_{\odot}$ diagram, which were 
calculated by the PARSEC code (Bressan et al. 2012, 2013) for $z=0.01$ 
(slightly metal-deficient case by $\sim 0.2$~dex lower than the solar metallicity) 
for the mass values from 0.8~$M_{\odot}$ to 4~$M_{\odot}$ with a step of 0.2~M$_{\odot}$.
The solid lines correspond to the shell-H-burning phase before 
He ignition, while the dashed lines to the He-burning phase after 
He ignition. Our sample stars are plotted by symbols (with the same meanings as in 
panel (a)). 
}
\end{figure*}

\setcounter{figure}{2}
\begin{figure*}[p]
  \begin{center}
    \FigureFile(120mm,160mm){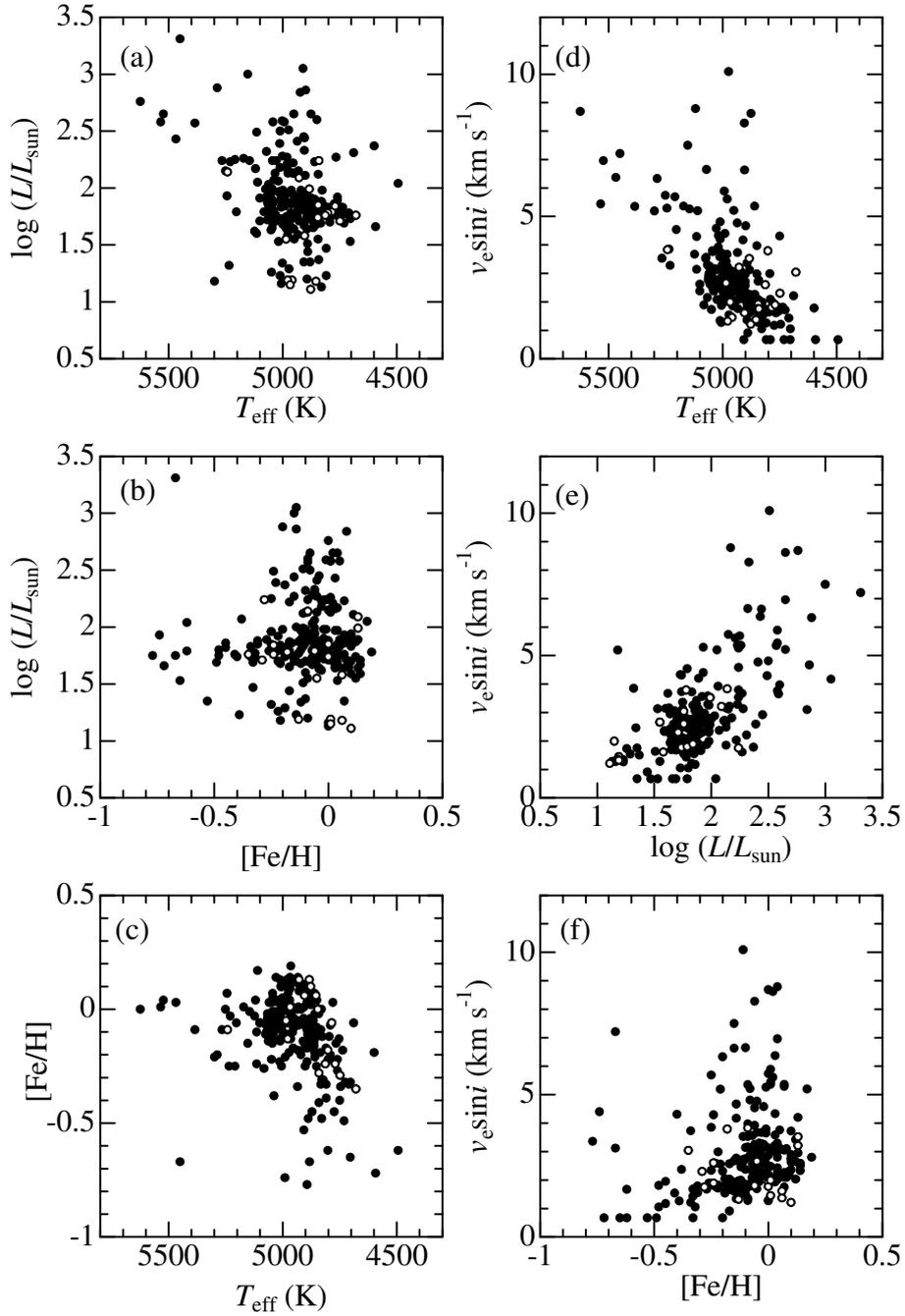}
  \end{center}
\caption{
Mutual correlations between the stellar parameters of 239 program stars.
(a) $\log L$ vs. $\log T_{\rm eff}$, (b) $\log L$ vs. [Fe/H],
(c) [Fe/H] vs. $T_{\rm eff}$, (d) $v_{\rm e}\sin i$ vs. $T_{\rm eff}$, 
(e) $v_{\rm e}\sin i$ vs. $\log L$, and (f) $v_{\rm e}\sin i$ vs. [Fe/H].
The 19 planet-host stars are denoted by open symbols in each panel;
we can see that they tend to have lower $v_{\rm e}\sin i$ and lower $\log L$,
but no specific trends are seen in terms of $T_{\rm eff}$ or [Fe/H].
}
\end{figure*}

\setcounter{figure}{3}
\begin{figure*}[p]
  \begin{center}
    \FigureFile(160mm,200mm){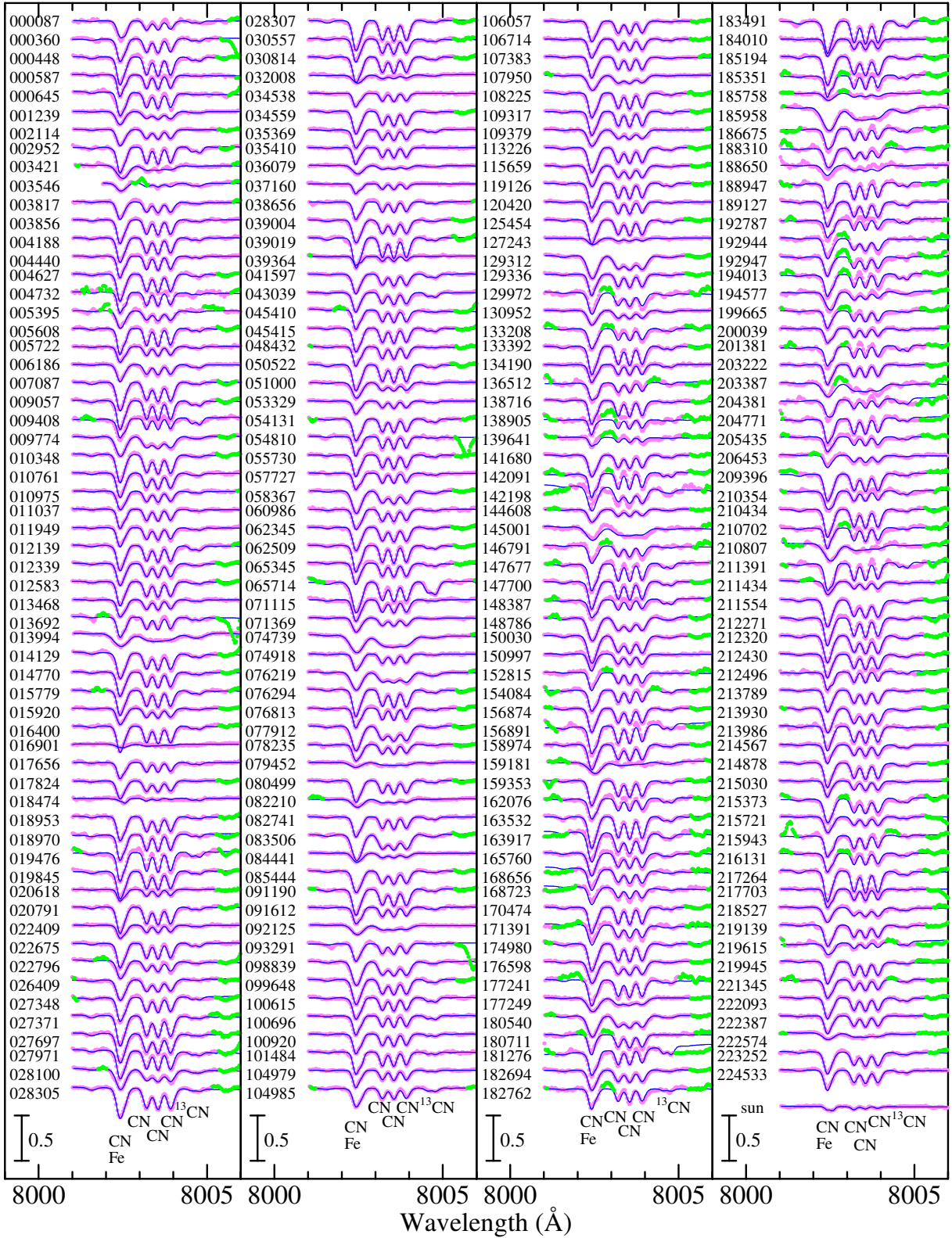}
  \end{center}
\caption{
Synthetic spectrum fitting in the 8001--8006~\AA\ region comprising 
Fe~{\sc i} and CN lines. The best-fit theoretical spectra
are shown by blue solid lines, and the observed data are plotted
by pink symbols (while those masked/disregarded in the fitting are 
highlighted in green). A vertical offset of 0.2 (in terms of the 
continuum-normalized flux) is applied to each 
relative to the adjacent ones. The spectra are arranged 
in the increasing order of HD number (indicated on the left to each 
spectrum), and the wavelength scale is adjusted to the laboratory frame
by correcting the radial velocity shift.
}
\end{figure*}

\setcounter{figure}{4}
\begin{figure*}[p]
  \begin{center}
    \FigureFile(120mm,180mm){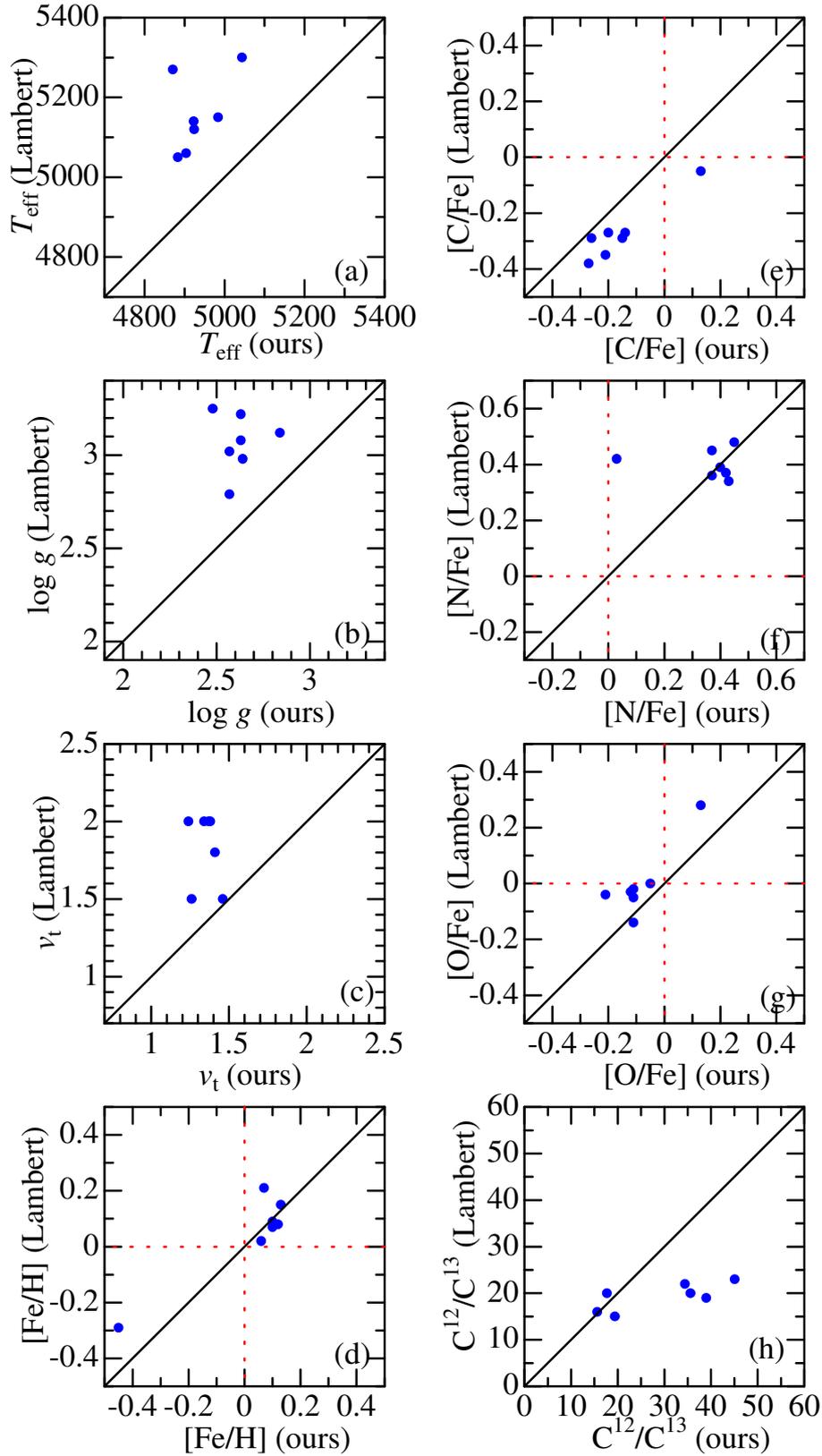}
  \end{center}
\caption{
Comparison of the adopted atmospheric parameters and the resulting 
abundances with those of Lambert and Ries (1981) for 7 stars in common. 
(a) $T_{\rm eff}$, (b) $\log g$, (c) $v_{\rm t}$, (d) [Fe/H]. (e) [C/Fe],
(f) [N/Fe], (g) [O/Fe], and (h) $^{12}$C/$^{13}$C.
}
\end{figure*}

\setcounter{figure}{5}
\begin{figure*}[p]
  \begin{center}
    \FigureFile(120mm,180mm){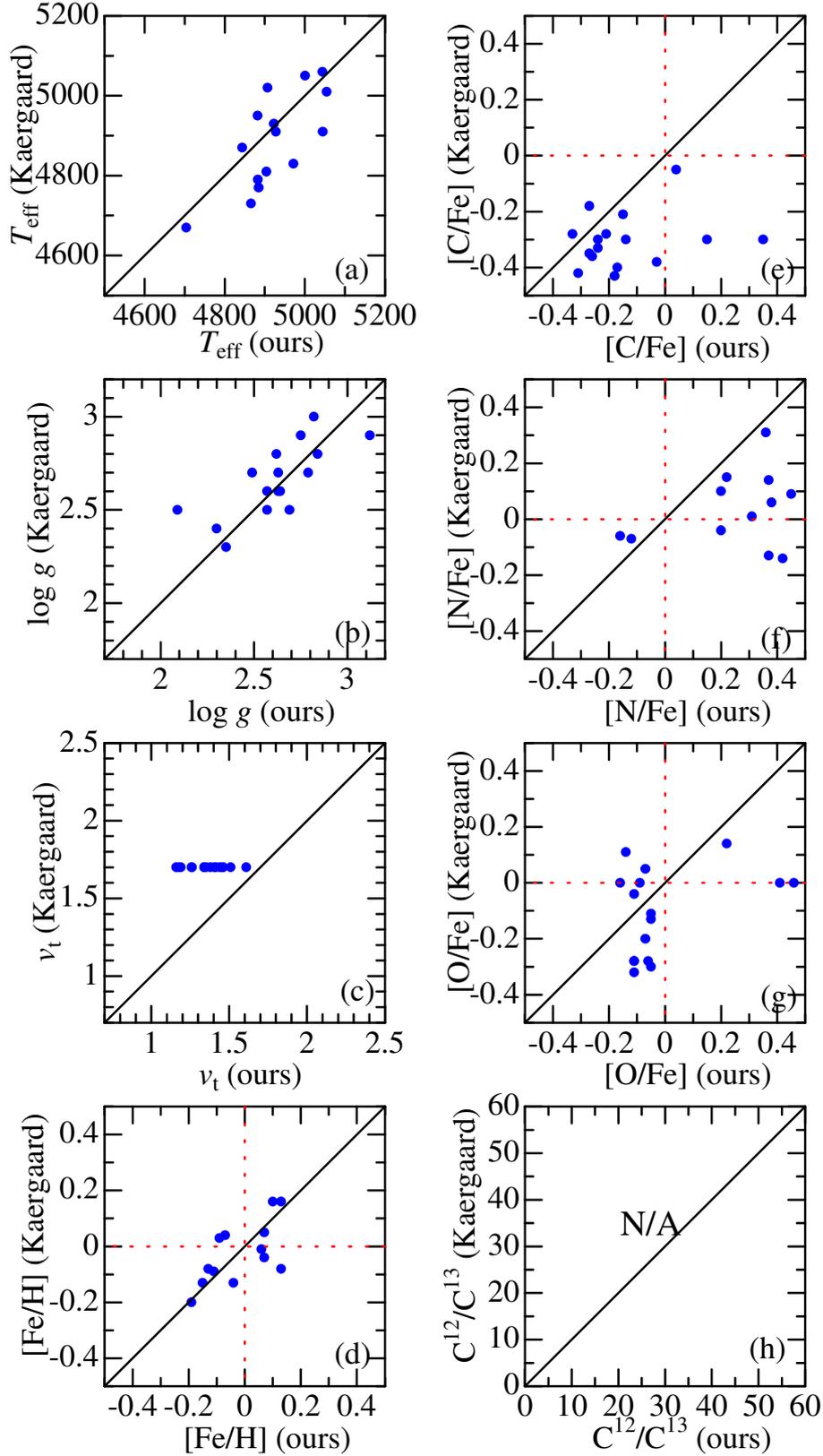}
  \end{center}
\caption{
Comparison of the adopted atmospheric parameters and the resulting 
abundances with those of K{\ae}rgaard et al. (1982) for 16 stars in common.
Otherwise, the same as in figure~5.
}
\end{figure*}

\setcounter{figure}{6}
\begin{figure*}[p]
  \begin{center}
    \FigureFile(120mm,180mm){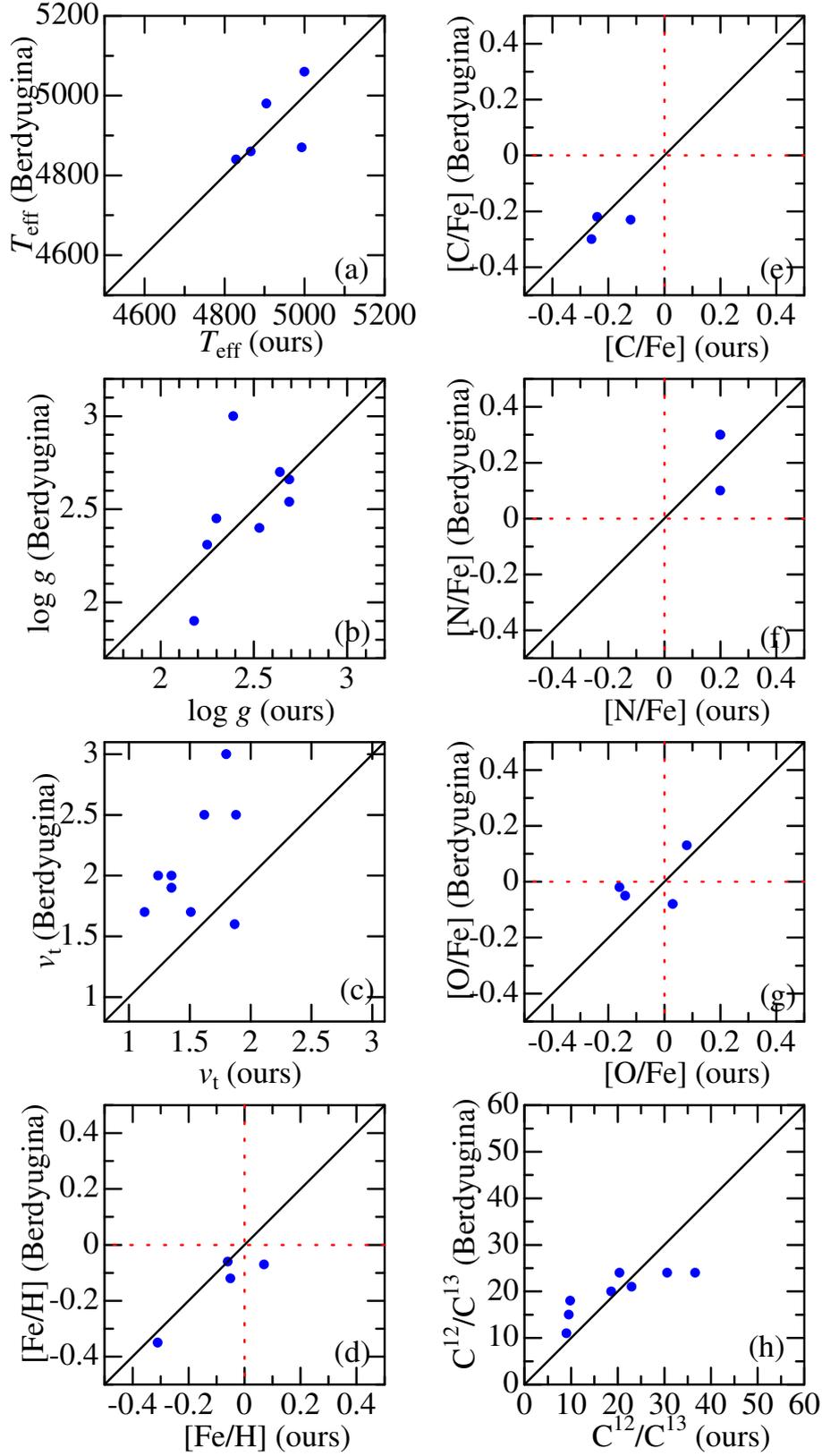}
  \end{center}
\caption{
Comparison of the adopted atmospheric parameters and the resulting 
abundances with those of Berdyugina (1993, 1994) for 9 stars in common.
Otherwise, the same as in figure~5. 
}
\end{figure*}

\setcounter{figure}{7}
\begin{figure*}[p]
  \begin{center}
    \FigureFile(120mm,180mm){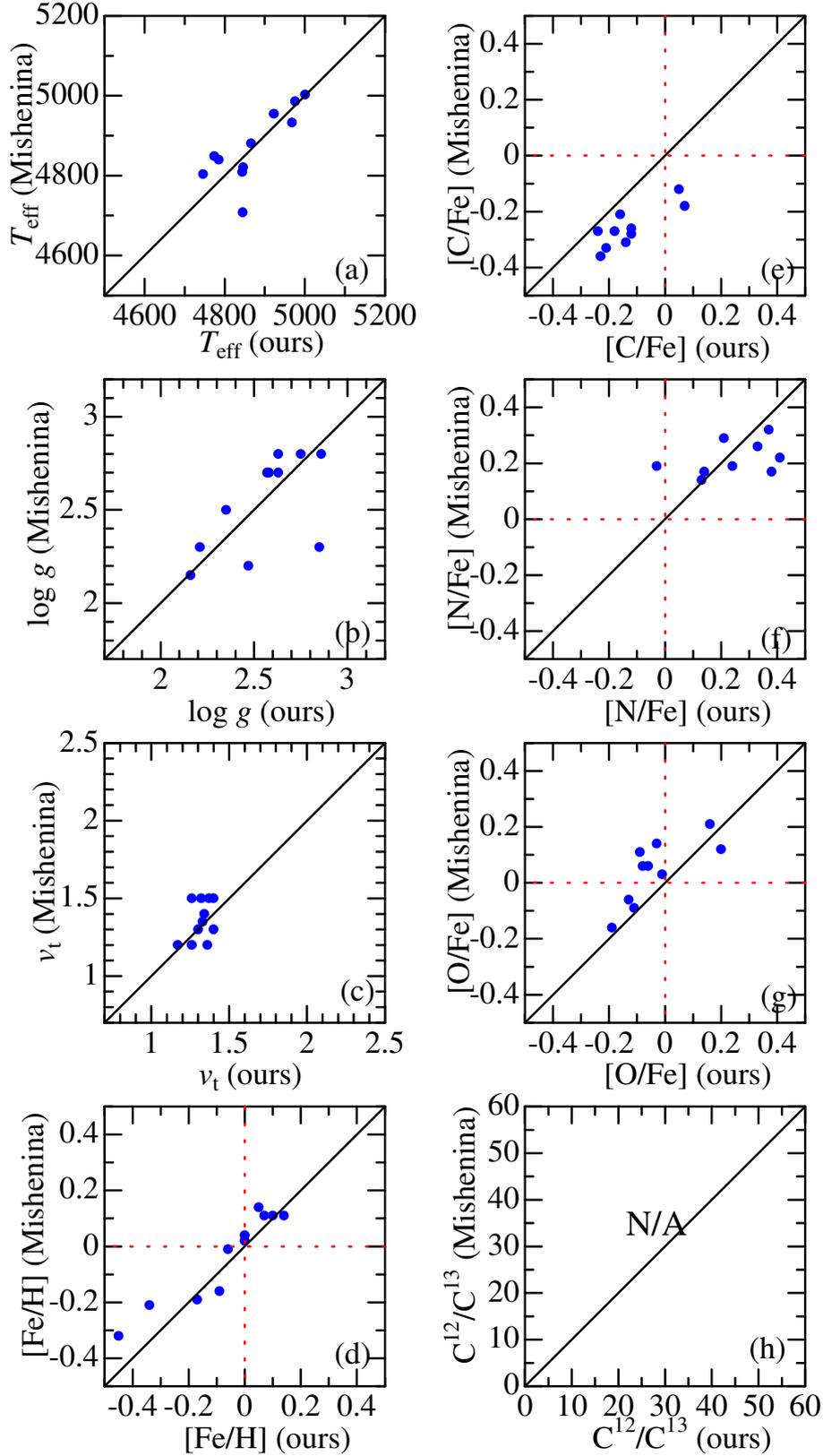}
  \end{center}
\caption{
Comparison of the adopted atmospheric parameters and the resulting 
abundances with those of Mishenina et al. (2006) for 11 stars in common. 
Otherwise, the same as in figure~5.
}
\end{figure*}

\setcounter{figure}{8}
\begin{figure*}[p]
  \begin{center}
    \FigureFile(120mm,180mm){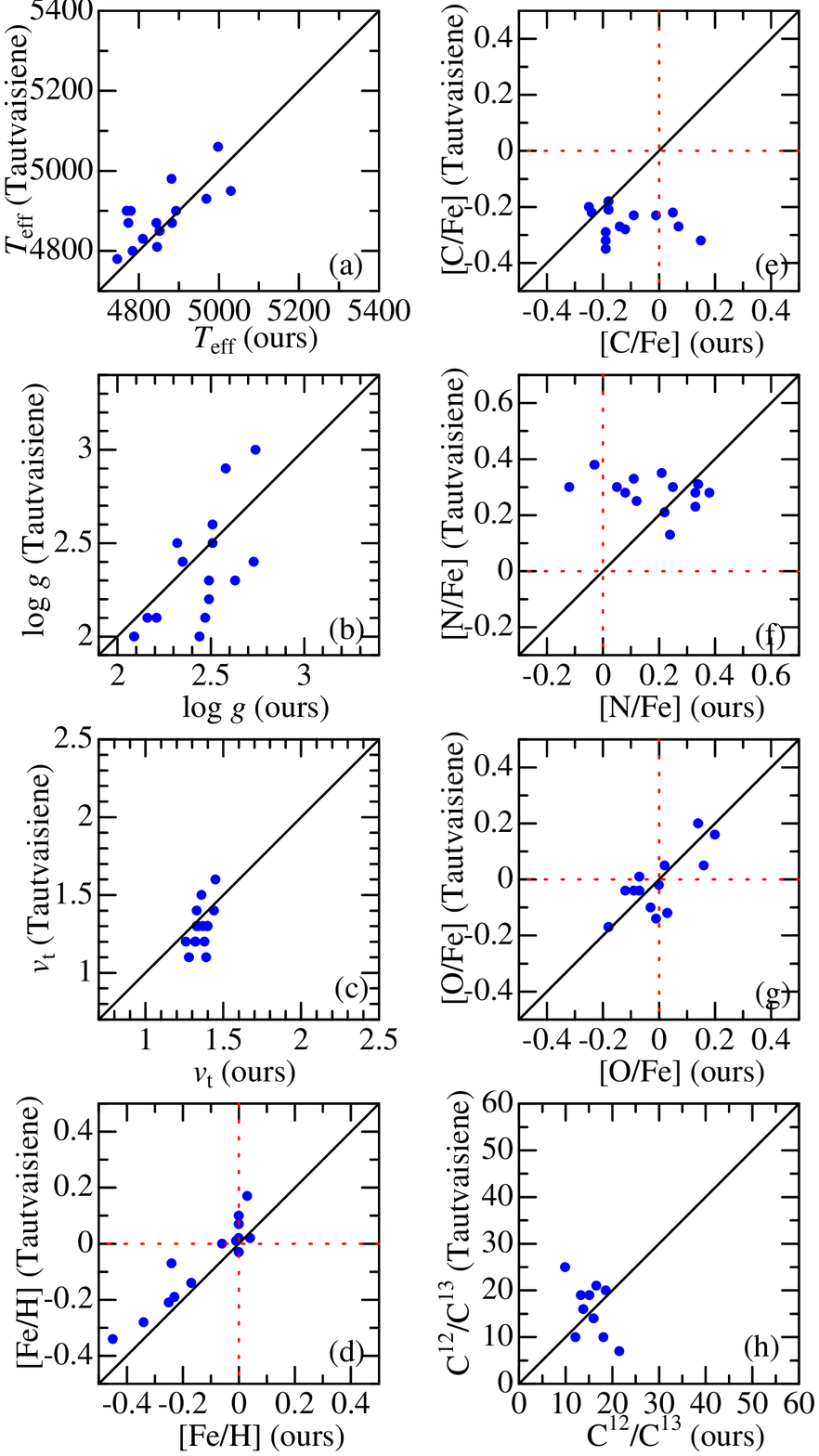}
  \end{center}
\caption{
Comparison of the adopted atmospheric parameters and the resulting 
abundances with those of Tautvai\u{s}ien$\dot{{\rm e}}$ et al. (2010, 2013) 
for 15 stars in common. Otherwise, the same as in figure~5.
}
\end{figure*}

\setcounter{figure}{9}
\begin{figure*}[p]
  \begin{center}
    \FigureFile(160mm,200mm){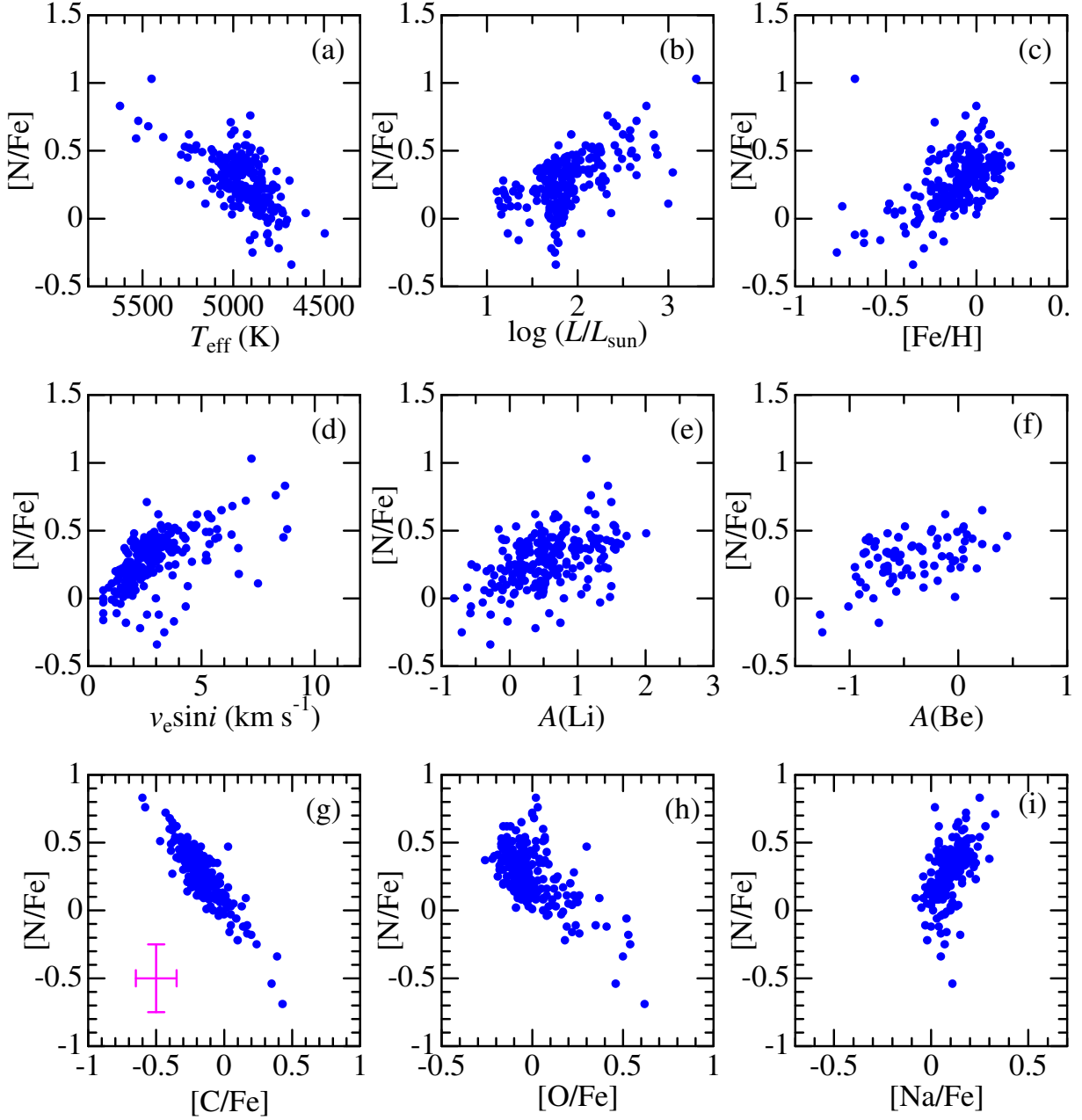}
  \end{center}
\caption{
[N/Fe] results plotted against stellar parameters 
and abundances of other elements. 
(a) $T_{\rm eff}$, (b) $\log L$, (c) [Fe/H],
(d) $v_{\rm e}\sin i$, (e) $A$(Li), (f) $A$(Be) (only reliable class-a values; 
cf. Takeda \& Tajitsu 2014), (g) [C/Fe], (h) [O/Fe], and (i) [Na/Fe]. 
In panel (g) are shown the typical error bars for [N/Fe] and [C/Fe] (cf. subsection~3.2)
}
\end{figure*}

\setcounter{figure}{10}
\begin{figure*}[p]
  \begin{center}
    \FigureFile(160mm,200mm){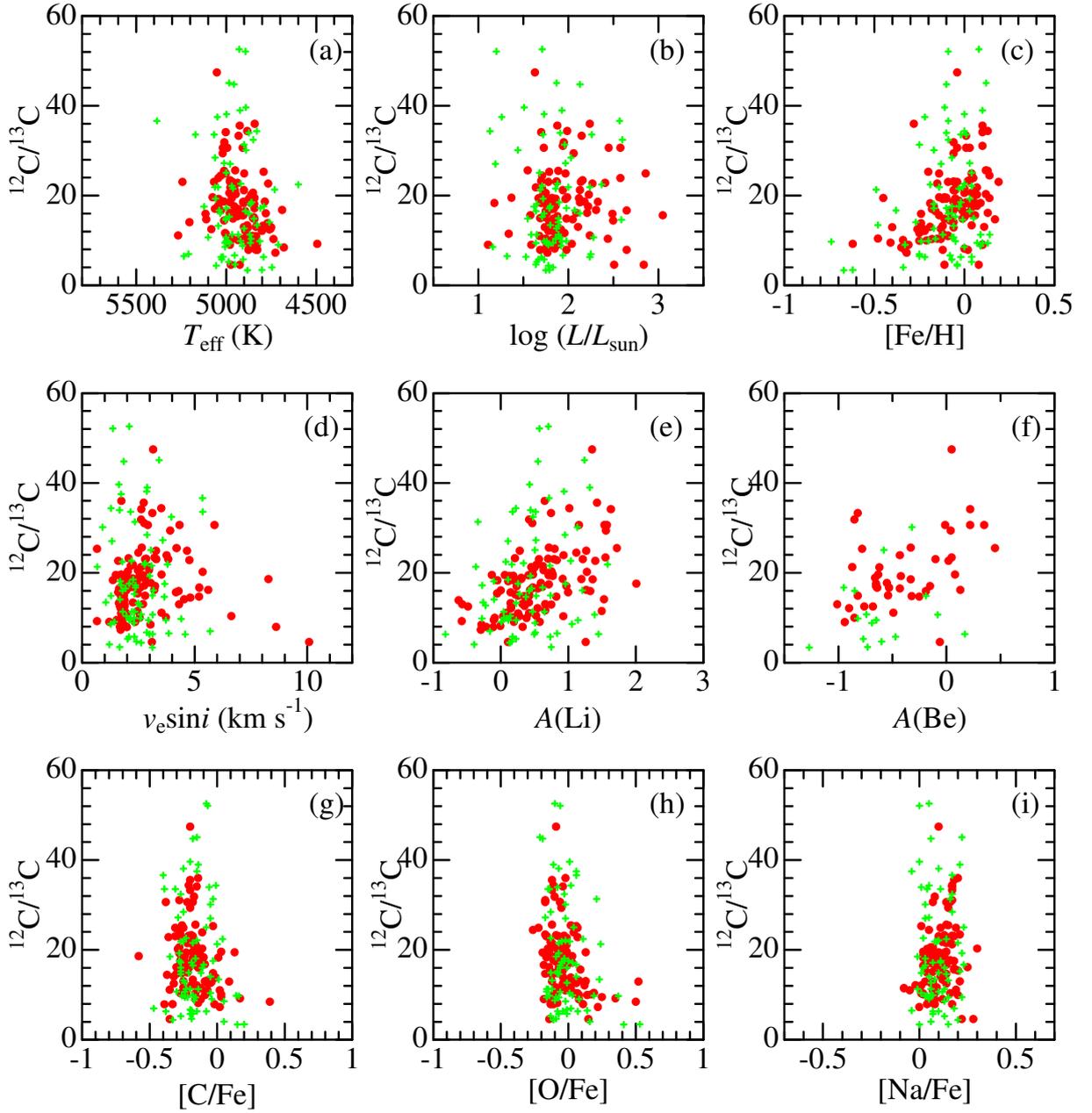}
  \end{center}
\caption{
$^{12}$C/$^{13}$C results plotted against stellar parameters 
and abundances of other elements. Filled circles and crosses
correspond to class-A (reliable) and class-B (less reliable) values,
respectively. Otherwise, the same as in figure~10.
}
\end{figure*}

\setcounter{figure}{11}
\begin{figure*}[p]
  \begin{center}
    \FigureFile(120mm,180mm){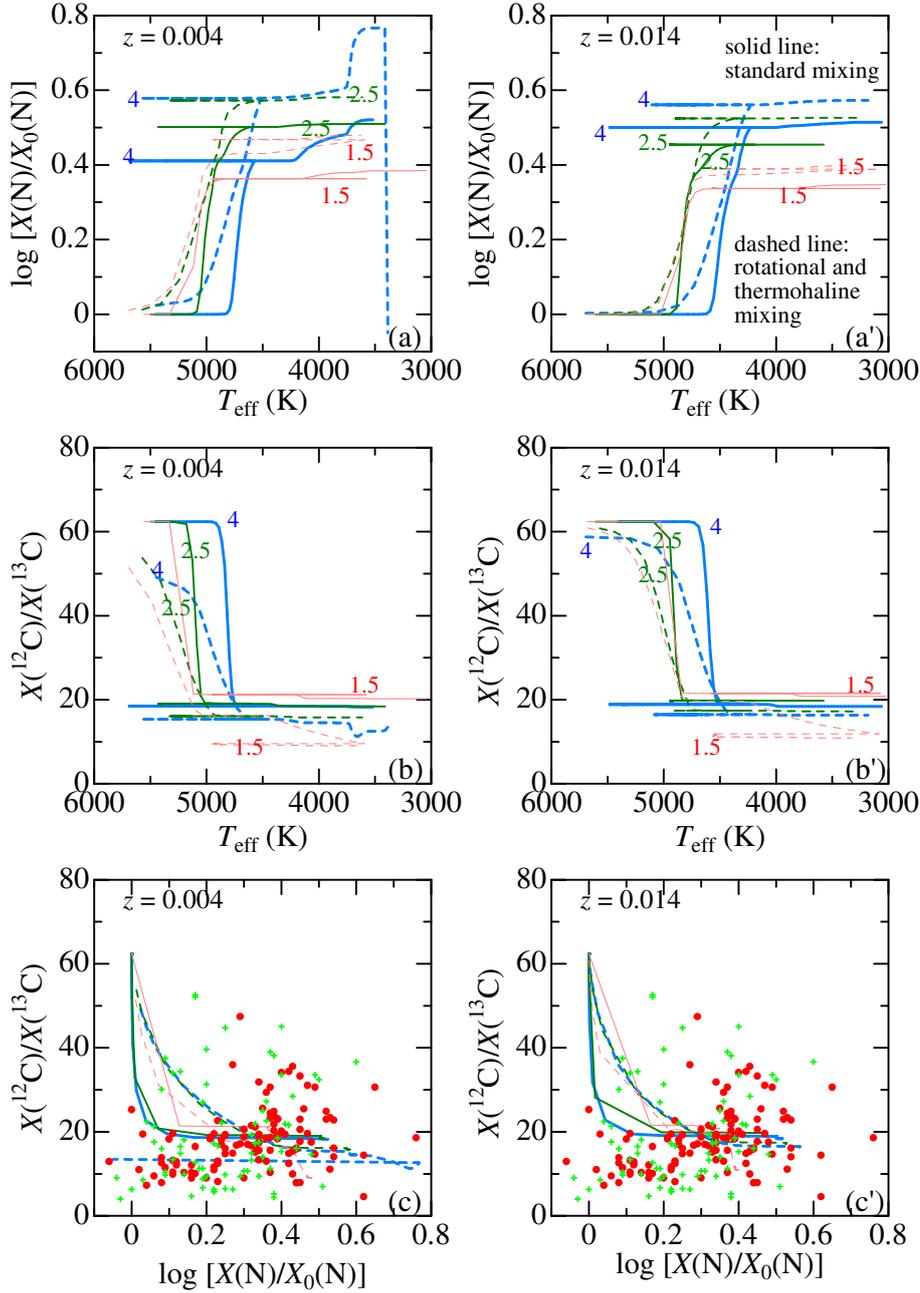}
  \end{center}
\caption{
$T_{\rm eff}$-dependence (top and middle panels) and mutual relation (bottom panels) 
of $\log [X({\rm N})/X({\rm N})_{0}]$ (logarithmic mass fraction ratio of N at 
the surface relative to the initial value) and $X(^{12}{\rm C})/X(^{13}{\rm C})$ ratio
theoretically simulated by Lagarde et al. (2012). 
The left panels are for $z = 0.004$ ($0.3 \times$ solar metallicity)
and the right are for  $z = 0.014$ ($1 \times$ solar metallicity).
The results corresponding to three stellar masses of 1.5, 2.5, 
and 4.0~$M_{\odot}$ are shown here, which are discriminated 
by line thickness (thin orange line, normal green line, and 
thick blue lines, respectively).
Different treatments of envelope mixing are discriminated by line types: 
standard treatment (solid line) and non-standard treatment including rotational 
and thermohaline mixing (dashed line). 
Here, we restricted the data only to those of the evolved red-giant stage 
satisfying the conditions of $T_{\rm eff} < 5700$~K and $age > 10^{7.5}$~yr.
In the bottom panels, our observed $^{12}$C/$^{13}$C data are also overplotted 
against [N/Fe] for comparison (see the caption of figure~11 for the meanings 
of the symbols).   
}
\end{figure*}

\setcounter{figure}{12}
\begin{figure*}[p]
  \begin{center}
    \FigureFile(120mm,160mm){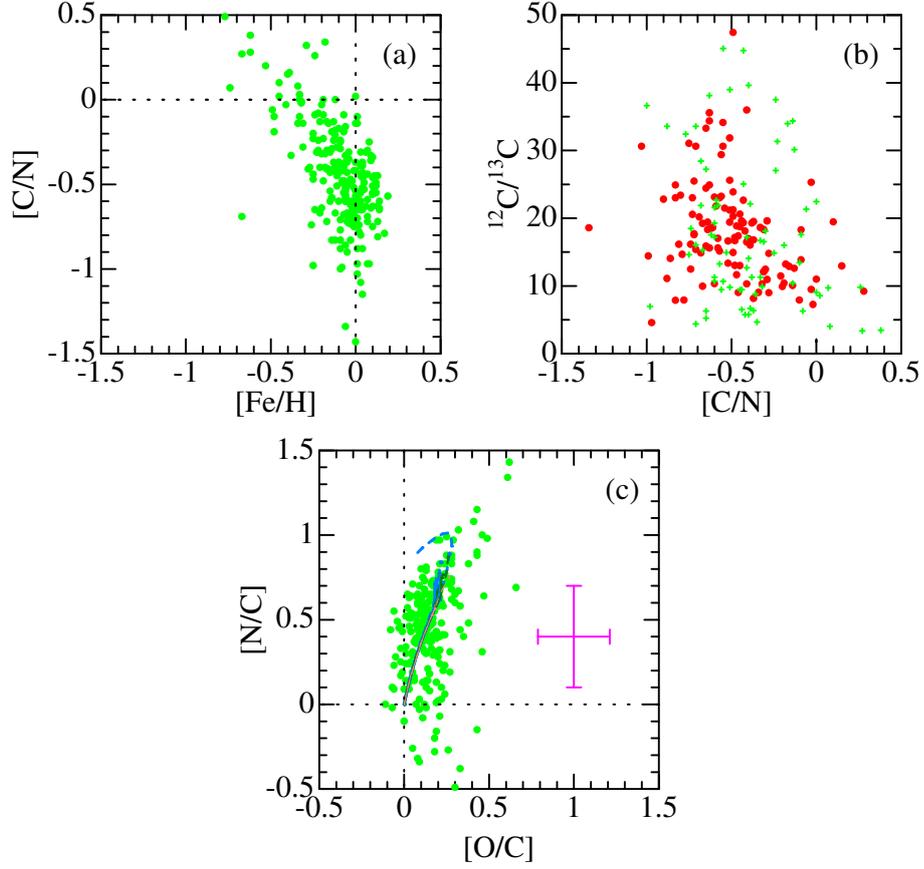}
  \end{center}
\caption{
(a) [C/N] vs. [Fe/H], (b) $^{12}$C/$^{13}$C vs. [C/N], and (c) [N/C] vs. [O/C] correlations 
derived for the 239 program stars. Panels (a) and (b) should be compared with Fig.~6 
and Fig.~11 of Lagarde et al. (2019), respectively; and panel (c) is for comparison
with Fig.~13h of Takeda, Jeong, and Han (2019). See the caption of figure~11 for
the meanings of the symbols in panel (b).
The typical error bars for [N/C] and [O/C] (cf. subsection~3.2) are depicted in panel (c).
In panel (c) are also shown the theoretically predicted relations computed by 
Lagarde et al. (2012) by lines, where 12 different loci  corresponding to 
the combination of three masses (1.5~$M_{\odot}$, 2.5~$M_{\odot}$, and 4~$M_{\odot}$), 
two different mixing treatments (standard mixing, non-standard mixing including 
thermohaline+rotational mixing), and two metallicities ($z = 0.004$ and $z = 0.014$) 
are overplotted in the same manner as in figure~12 (though hardly discernible 
from each other).
}
\end{figure*}

\end{document}